\documentclass[12pt,preprint]{aastex}

\shorttitle{Estimation of Microlensing Magnifications}
\shortauthors{Albrow}

\begin{document}

\title{Early Estimation of Microlensing Event Magnifications}

\author{Michael D. Albrow}
\affil{Department of Physics and Astronomy, University of Canterbury,
    Private Bag 4800, Christchurch, New Zealand}
\email{Michael.Albrow@canterbury.ac.nz}

\begin{abstract}
Gravitational microlensing events with high peak magnifications
provide a much enhanced sensitivity
to the detection of planets around the lens star. However, 
estimates of peak 
magnification during the early stages of an event by 
means of $\chi^2$ minimization frequently involve an 
overprediction, making observing campaigns with
strategies that rely on these predictions inefficient.

I show that a rudimentary Bayesian formulation, 
incorporating the known statistical characteristics
of a detection system, produces much more accurate predictions
of peak magnification than $\chi^2$ minimisation. 
Implementation of this system
will allow efficient follow-up observing programs that
focus solely on events that contribute to planetary
abundance statistics.
\end{abstract}


\section{Introduction}

Following the suggestion of \citet{Paczynski86}, several
collaborations, notably MACHO and EROS, began to search for
gravitational microlensing towards the Magellanic Clouds as an
indicator of compact objects in the halo of the Milky Way
\citep{Alcock93, Aubourg93}. At about the same time, the OGLE 
collaboration began a survey in the direction of the Galactic
bulge \citep{Udalski92,Udalski93}.
It was soon found that a much higher event rate occurred
in fields towards the Galactic bulge relative to the rate
towards the Magellanic Clouds \citep{Udalski94a,
Alcock95,Alcock97a}.
Since 1990, approximately
1000 such events have been detected \citep{Alcock00,Udalski00}.

Several groups including PLANET (Probing Lensing Anomalies NETwork,
\citealt{Albrow98,Albrow01,Dominik02,Gaudi02}),
MPS (Microlensing Planet Search, \citealt{Rhie99}) and $\mu$FUN
(Microlensing Follow-Up Network, \citealt{Yoo03}) monitor events much
more intensively than the survey groups in order to identify anomalous
behavior that can signal the presence of a planet associated with the
lens star. 
High-magnification events in particular (those with $A_{\rm 0} \gtrsim 10$) 
attract the attention of follow-up groups
since it is these that are most likely to give
detectable planetary signals
\citep{Griest98,Gaudi98}. In addition, for high magnification events the
angular size of the source star may be non-negligible in comparison to the 
lens-source angular separation. In these cases the lightcurves of the events
can provide the possibility
to determine the lens-source relative proper motion
\citep{Gould94,Alcock97b}
and atmospheric properties of the source \citep{Heyrovsky03}.

In the first years of operation, when microlensing alerts came
primarily from the MACHO collaboration, detected event rates were low
enough that PLANET could monitor almost all potentially interesting
events with ease. For the last two years (the 2002 and 2003 Bulge
seasons), this has not been the case, due to the much 
improved alert
rate since the advent of the OGLE III early warning system (EWS),
\url{http://www.astrouw.edu.pl/~ogle/ogle3/ews/ews.html}
\citep{Udalski94b,Udalski03}. In excess of 400 events were alerted by 
the EWS in
each of these years. In addition, approximately 75 events were alerted
in 2003 by the MOA collaboration \citep{Bond02} although some of these
were duplicates of EWS events.  We are now in an era in which a careful
selection of events is necessary to optimize planet detection and
exclusion productivity. For this reason, follow-up groups require
accurate predictions of eventual maximum amplications in the early
days following a detection. For the remainder of this paper I will
focus exclusively on events detected by the OGLE III EWS.
 
\section{Fitting microlensing lightcurves}

Most microlensing events are well fitted by a point-source
point-mass-lens (PSPL) model for the magnification $A(t)$ at time $t$,
\begin{eqnarray}
A(u)    &  = & \frac{u^2+2}{u \sqrt{u^2+4}} , \\
u(\tau) &  = & \sqrt{u_{\rm 0}^2+\tau^2} , \\
\tau(t) &  = & \frac{t-t_{\rm 0}}{t_{\rm E}} .
\end{eqnarray}
The impact parameter $u$ is the angular separation
between the  source and lens
measured in units of the angular Einstein radius,
\begin{equation}
\theta_{\rm E} = \sqrt{\frac{4GMD_{\rm LS}}{c^2 D_{\rm L} D_{\rm S}}},
\end{equation}
where $M$ is the mass of the lens, $D_{\rm S}$ is the observer-source distance, $D_{\rm L}$ the 
observer-lens distance, $D_{\rm LS}$ the source-lens distance and
$u_{\rm 0}$ is the impact parameter at $t_{\rm 0}$, the time of maximum 
magnification. 
The Einstein radius crossing time, 
\begin{equation}
t_{\rm E} = \frac{\theta_{\rm E}}{\mu_{\rm rel}},
\end{equation}
where $\mu_{\rm rel}$ is the relative proper motion between the
lens and source.

The lightcurve of a PSPL
event can thus be characterised by $3+2n$ parameters, ($t_{\rm 0},
u_{\rm 0}, t_{\rm E}$) plus for each $n$ telescope + filter combinations,
the unmagnified (baseline)
magnitude of the source star $m_{\rm base}$ and the blending parameter
$f_{\rm bl}$,
where $1 - f_{\rm bl}$ is the fraction of blended (non-lensed) light. 
The maximum magnification,
$A_{\rm 0} = A(u_{\rm 0})$ frequently replaces $u_{0}$ as a parameter.

The conventional method for predicting peak magnifications is to use
$\chi^2$ minimization techniques to fit PSPL models to data from 
OGLE III (possibly supplemented by a follow-up group's own data) as they
become available. Such fits are continuously updated and the subsequent 
predictions revised as data accumulate. Experience has shown that
early predictions of eventual maximum magnification using these methods
systematically yield overpredictions, strongly limiting the 
usefulness of such estimates.
In particular, very large maximum magnifications (with large uncertainties)
are often predicted for events that turn out to be of rather low
amplitude. Valuable observing time is often wasted monitoring such events
in order to confirm their nature.

The reason for this overprediction of values for $A_{\rm 0}$ is that
in using $\chi^2$ minimization for a predictive purpose, one
implicitly assumes that all parameter values are equally
likely. However, for microlensing events this is far from being the
case. From a purely geometrical perspective, high magnification events
are exceedingly rare. In practice, being of high magnification, they
have a higher probability of detection by a survey group. It is the
individual detection efficiency of a survey convolved with the
intrinsic event rates (both of these as a function of event
parameters) that determines the magnification probability, given that
an event has been detected.

These ideas can be given a quantitative basis in a Bayesian
formulation of the problem. The merits of the Bayesian approach to
statistical analysis have been discussed at length elsewhere and will
not be reargued \citep{Loredo90,Sivia96}. Here we simply note that a
Bayesian formulation with appropriate priors should produce an
unbiased estimate of the eventual microlensing event parameters during
the rising part of a lightcurve.

From Bayes' theorem, the probability density for a microlensing event
to have a certain set of parameter values ${\underline \theta}$, given
the fact, $O$, that it has been detected by the OGLE III EWS and that
data $D$ have been acquired,
\begin{equation}
p({\underline \theta}|D,O) \propto 
          p(D|{\underline \theta, O}) \; p({\underline \theta}|O),
\end{equation}
where ${\underline \theta} = (A_{\rm 0},t_{\rm E},t_{\rm 0},m^{\rm j}_{\rm base},
f^{\rm j}_{\rm bl},j=1..n)$. 
The second term on the right hand side of 
equation (6), $p({\underline
\theta}|O)$ (known as the prior), is the underlying probability
density for ${\underline \theta}$ given a detection, i.e.
$p({\underline \theta}|O) d{\underline \theta}$ is the probability 
that ${\underline \theta}$ is in the range 
$[{\underline \theta},{\underline \theta}+d{\underline\theta}]$. 
It is this
function that incorporates both the natural event occurrence
probability and the particular parameter sensitivities of the
detection system. The first term, $p(D|{\underline \theta},O) = {\cal
L}({\underline \theta}|D)$, the likelihood function for ${\underline
\theta}$ given $D$. Data from the OGLE III survey consist of
$I$-band magnitudes and their uncertainties $(m_{\rm i},\sigma_{\rm i})$ at
time $t_{\rm i}$ (i.e. $n=1$).  
I assume each $m_{i}$ to be drawn randomly from a
normal distribution $N(m_{\rm i,0},\sigma_{\rm i})$ where the true
value of the magnitude at time $t_{\rm i}$ is $m_{\rm i,0}$.  This
implies that
\begin{equation}
{\cal L}({\underline \theta}|D) \propto e^{-\chi^2/2}
\end{equation}
where
\begin{equation}
\chi^2 = \sum_{i} \left( \frac{m_{i}-m(t_{\rm i},{\underline \theta})}
                         {\sigma_{i}} \right)^2,
\end{equation}
and $m(t_{\rm i},{\underline \theta})$ is the magnitude evaluated from
the model parameterised by ${\underline \theta}$.

Analagous to a $\chi^2$ minimization, the value of ${\underline
\theta}$ that maximises $p({\underline \theta}|D,O)$ (i.e. the posterior
mode) is taken as the
best estimator of ${\underline \theta_{\rm 0}}$, the true value of
${\underline \theta}$. In the absence of a prior, this solution
reduces to the minimum $\chi^2$ solution. We stress here that when
sufficient data are available to constrain a fit to a certain event,
e.g. when the event is over, the Bayesian and $\chi^2$ minimization
techniques give the same parameter values and the choice of prior is
largely irrelevant. In other words, the solution is not driven by
prior probabilities when sufficent empirical information is available
(see \citealt{Sivia96}, Chapter 2 for a discussion of this point).

If the parameters ${\underline \theta}$ are statistically independent
quantities\footnote{ This is not necessarily the case given that a
detection system may preferentially select events with parameter
corelations, however inspection of EWS-detected events has not revealed
any such corelations as yet.}, $p({\underline \theta}|O)$ factorizes as
\begin{equation}
p({\underline \theta}|O) = p(A_{\rm 0}) \; p(t_{\rm E}) \; p(t_{\rm 0}) 
                        \; p(m_{\rm b}) \; p(f_{\rm bl}),
\end{equation}
where for brevity I have omitted "$|O$" in the probability densities
on the right hand side of
the equation.
It is often more convenient to work in decadic logarithmic units for
several of these quantities, in which case
\begin{equation}
p({\underline \theta}|O) = \frac{p(\lg A_{\rm 0}) \; p(\lg(t_{\rm E}/t^{*}))
       \; p(\lg(\Delta t_{\rm 0}/t^{*})) \; p(m_{\rm b})p(f_{\rm bl})}
           {(\ln 10)^{3} \; A_{\rm 0} \; (t_{\rm E}/t^{*}) \;
           (\Delta t_{\rm 0}/t^{*})} ,
\end{equation}
with $t^{*}$ being an arbitrary unit of time.
Here I define $\Delta t_{\rm 0}$ to be the time
to peak magnification from an initial ``alert date''.
For the remainder of this paper I adopt $t^{*} = 1\;{\rm d}$.

It is worth noting that even if each parameter in ${\underline \theta}$
is independent in $p({\underline \theta}|O)$, they are not independent
in the 
likelihood function $p(D|{\underline \theta},O)$ and hence
not independent in $p({\underline \theta}|D,O)$. Thus
when fitting a model to a lightcurve, particularly when only early 
data are available, the fitted maximum magnification is affected
not only by the prior on $A_{\rm 0}$ but also by the priors on the other
other parameters.

\subsection{Inclusion of a blending parameter}

The criterion used by the OGLE III EWS 
is that a blending parameter is only used when it is more than 3-$\sigma$
less than unity and is larger than its formal uncertainty.
In this paper I use the odds ratio test, 
a natural way to decide between two different models. I define the 
odds ratio
\begin{equation}
\label{OReqn1}
\frac{p(\tilde{\underline \theta}|D,O)}{p({\underline \theta}|D,O)} =
\frac{p(D|\tilde{\underline \theta},O)}{p(D|{\underline \theta},O)} 
\frac{p(\tilde{\underline \theta}|O)}{p({\underline \theta}|O)},
\end{equation}
where $\tilde{\underline \theta} = (A_{\rm 0},t_{\rm E},t_{\rm 0},m^{\rm j}_{\rm base},j=1..n)$ 
indicates the set of model parameters without blending
($n=1$ when considering only EWS data).
Having no {\it a priori} indication about whether to include blending I choose
$p(\tilde{\underline \theta}|O) = p({\underline \theta}|O)$. 
If we assume a unform prior probability density
for $f_{\rm bl}$ in the range $0 < f_{\rm bl} \leq 1$ and zero
outside this range, and assuming a Gaussian probability density
function for $f_{\rm bl}$ about $f_{\rm bl,0}$, it can be shown (see
for instance \citealt{Sivia96} Ch 4) that equation (\ref{OReqn1}) reduces to
\begin{equation}
\frac{p(\tilde{\underline \theta}|D,O)}{p({\underline \theta}|D,O)} =
\frac{p(D|\tilde{\underline \theta}_{\rm 0})}{p(D|{\underline \theta}_{\rm 0})} 
\frac{1}{\sqrt{2 \pi} \sigma_{f_{\rm bl}}}
\end{equation}
for cases in which $f_{\rm bl,0}$ is more than several $\sigma_{f_{\rm bl}}$
away from the cutoffs imposed by the prior. Otherwise, for $f_{\rm bl,0}$
close to 1, equation (\ref{OReqn1}) becomes
\begin{equation}
\frac{p(\tilde{\underline \theta}|D,O)}{p({\underline \theta}|D,O)} =
\frac{p(D|\tilde{\underline \theta}_{\rm 0})}
     {p(D|{\underline \theta}_{\rm 0})} 
\frac{1}{\sqrt{\frac{\pi}{2}} \sigma_{f_{\rm bl}} 
\left(1 + {\rm erf}\left(\frac{1-f_{\rm bl,0}}{\sqrt{2} \sigma_{f_{\rm bl}}} \right) \right) } ,
\end{equation}
while for $f_{\rm bl,0}$ close to 0
\begin{equation}
\frac{p(\tilde{\underline \theta}|D,O)}{p({\underline \theta}|D,O)} =
\frac{p(D|\tilde{\underline \theta}_{\rm 0})}
     {p(D|{\underline \theta}_{\rm 0})} 
\frac{1}{\sqrt{\frac{\pi}{2}} \sigma_{f_{\rm bl}} 
\left(1 + {\rm erf}\left(\frac{f_{\rm bl,0}}{\sqrt{2} \sigma_{f_{\rm bl}}} \right) \right) } .
\end{equation}
The odds ratio is then made up of two terms. The first of these
represents a relative ``goodness of fit'' between the two models 
while the second is the ``Occam penalty'' for introducing a new
parameter. Only when $f_{\rm bl,0} < 1$
and the odds ratio is less than unity is a blending parameter used in this 
analysis.

\section{Statistical properties of the 2002 OGLE events}

I have used the set of microlensing events detected by
the EWS in 2002 to determine the parameter priors. Since our interest
is in the set of PSPL events, I have removed 41 events that showed
deviations from PSPL behavior from this analysis. Excluded events were
numbered 
18, 23, 40, 51, 68, 69, 77, 80, 81, 99, 113, 119, 126, 127, 128, 129, 131, 135, 
143, 149, 159, 175, 194, 202, 203, 205, 215, 228, 229, 232, 238, 254, 255, 256, 266, 273,
307, 315, 339, 348, 360, out of the complete set of 389 alerts. 

Parameter values have been obtained for these events using a simplex
downhill method to minimise $\chi^2$ (EWS estimates of $A_{\rm 0}$,
$t_{\rm E}$, $m_{\rm b}$, $f_{\rm bl}$ can also be obtained from the
EWS web page).
For $\Delta
t_{\rm 0}$ we require an objective definition of an ``alert date''
that can be applied to all events. I have arbitrarily chosen a working
definition of an alert date as being the date at which three
successive data points have been more than 1-$\sigma$ brighter than
$m_{\rm b}$, the baseline magnitude.  In practice, $m_{\rm b}$ can be
determined separately from and in advance of the other parameters.

\begin{figure}
\plotone{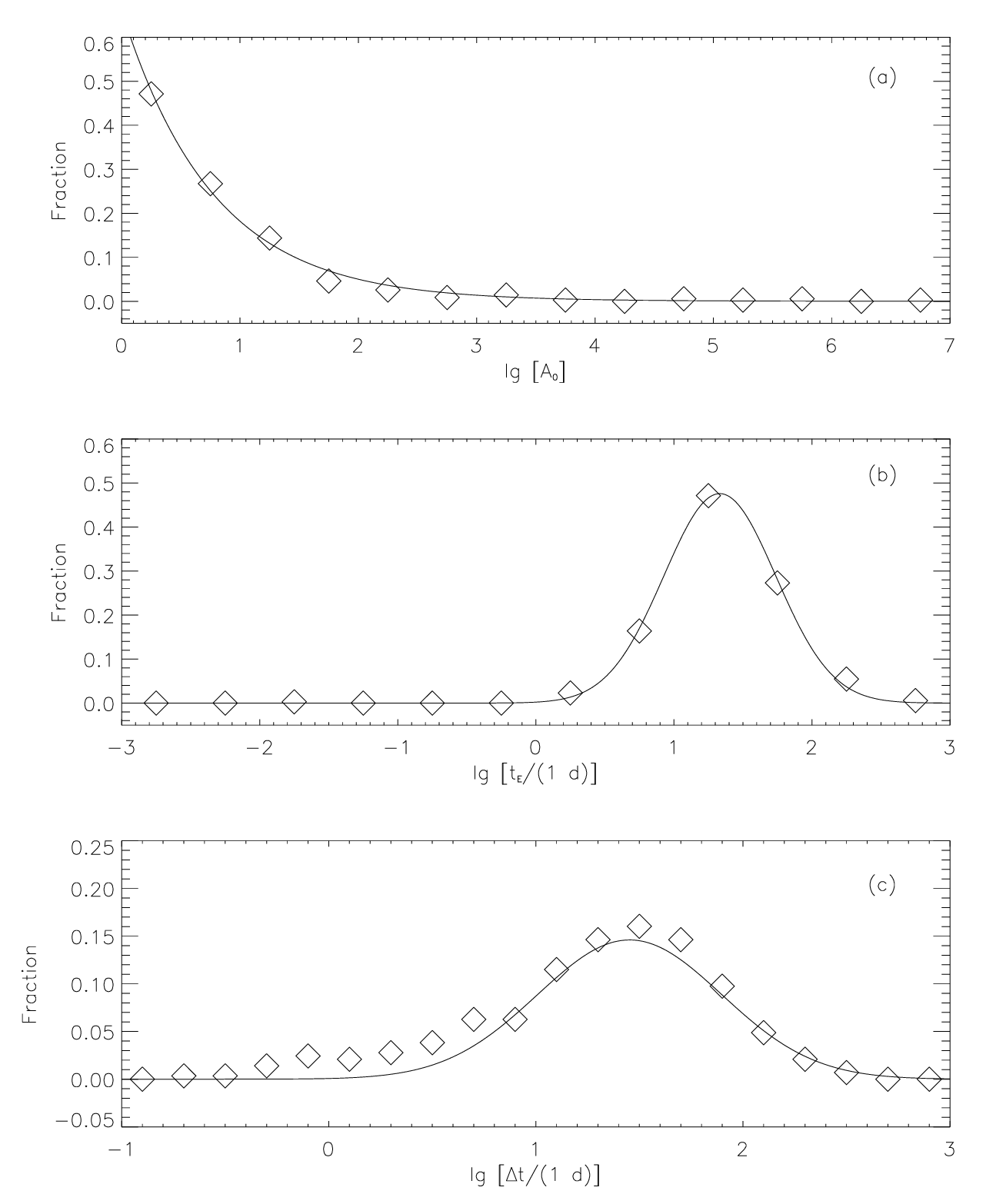}
\caption{Distributions of (a) peak magnifications 
         (b) Einstein timescales and (c) times from alert
         to peak magnification
         from my own fits to the 2002 OGLE event data. 
         The solid lines are the adopted
         Bayesian priors based on these distributions. \label{fig1}}
\end{figure}

The distributions of $\lg A_{\rm 0}$, $\lg t_{\rm E}$ and 
$\lg \Delta t_{\rm 0}$ are shown
in Figure~\ref{fig1}.
For the purposes of obtaining Bayesian prior probability densities for these
quantities, the distribution functions are adequately represented 
by the following empirically-chosen functions, also shown in Figure~\ref{fig1}:
\begin{eqnarray}
p(\lg A_{\rm 0}) & = & 0.660 \exp \left[ -1.289 \lg A_{\rm 0} \right]  \\
p(\lg (t_{\rm E}/t^{*})) & = & 0.476 \exp \left[ -(\lg (t_{\rm E}/t^{*}) - 1.333)^2 /0.330 \right]  \\
p(\lg (\Delta t_{\rm 0}/t^{*})) & = & 0.156 \exp \left[ -(\lg (\Delta t_{\rm 0}/t^{*}) - 1.432)^2 /0.458 \right].
\end{eqnarray}

\begin{figure}
\plotone{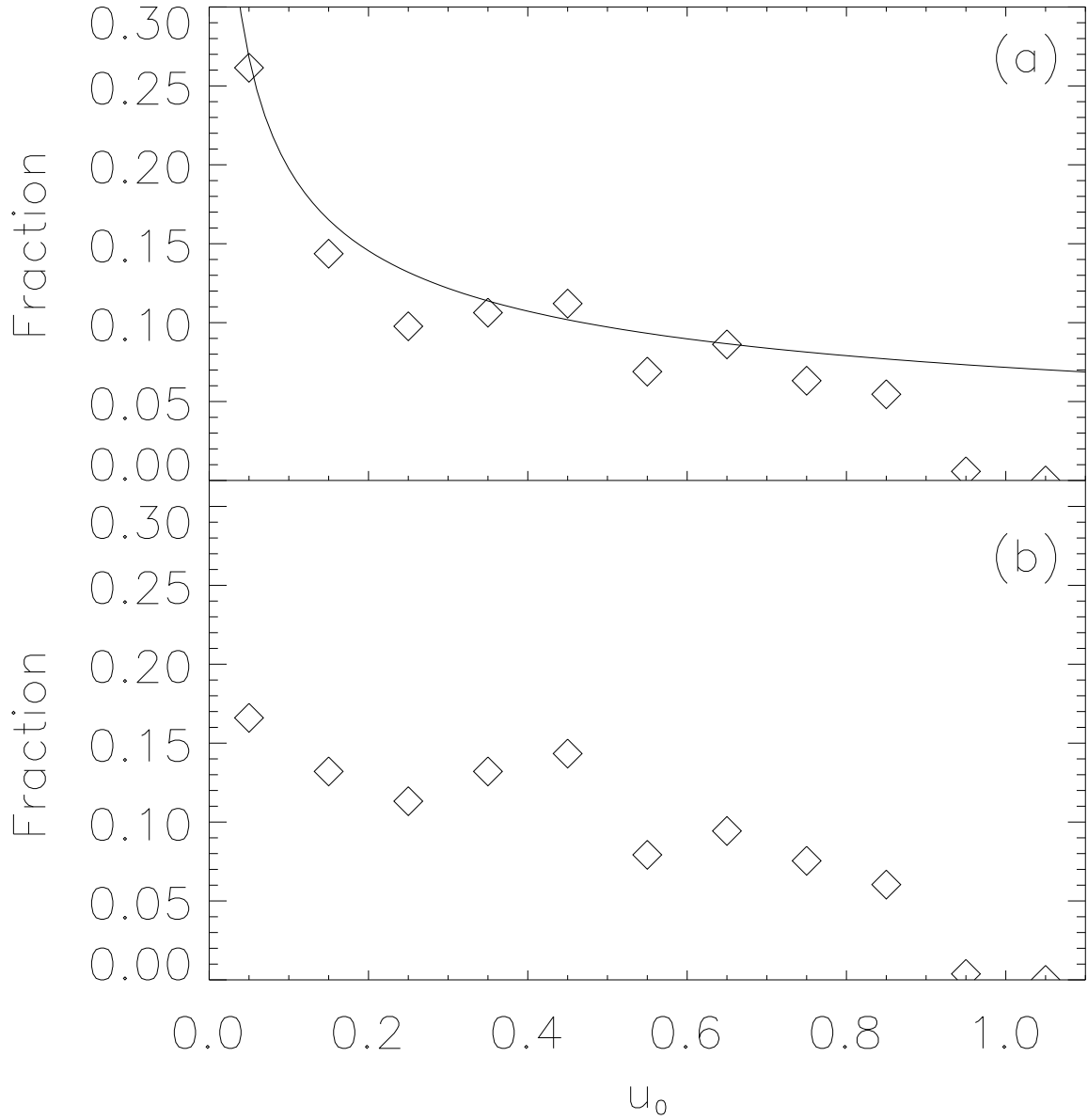}
\caption{Fractional distribution of $u_{\rm 0}$ for the 2002 OGLE
         PSPL events. Panel (a) includes the entire data set while
         panel (b) excludes events for which the uncertainty in $A_{\rm 0}$
         is greater than 50\%. The solid line in panel (a) represents the
         Bayesian prior on $\lg A_{\rm 0}$ transformed to $u_{\rm 0}$ 
         and scaled to the first data point. \label{fig2}}
\end{figure}

It is also instructive to examine the distribution of $u_{\rm 0}$,
shown in Figure~\ref{fig2}(a). In the absence of any selection effects,
this distribution should be uniform. In fact, there is an enhanced
sensitivity to detection of high magnification (low $u_{\rm 0}$) 
events and a rapid decrease in sensitivity for $u_{\rm 0} \gtrsim 0.85$.
Figure ~\ref{fig2}(a) also shows the shape of the adopted prior on 
$\lg A_{\rm 0}$ (eq.~15) when transformed to $u_{\rm 0}$.

Figure~\ref{fig2}(b) shows the same data but excluding those events
where $A_{\rm 0}$ has a formal uncertainty greater than 50\%. This
illustrates that many high amplification events have maxima that are
poorly constrained from OGLE data alone.

\section{Application to 2003 OGLE alerts}

As a test of the Bayesian method, I have applied a fitting procedure that
maximises $p({\underline \theta}|D,O)$ to a sample of the
PSPL events alerted in real time by the
OGLE III EWS in 2003. These consist of events OGLE-2003-BUL-138 to
OGLE-2003-BUL-462 and excluding events numbered 145, 160, 168, 170,
176, 192, 200, 230, 236, 252, 260, 266, 267, 271, 282, 286, 293, 303,
306, 311, 359, 380, 419 that do not appear to be due to PSPL microlensing
and 188, 197, 245, 263, 274, 297, 387, 399, 407, 412, 413, 417, 420, 422, 429,
430, 432, 433, 435, 437, 440, 441, 442, 443, 444, 449, 450, 452, 453, 454, 455,
457, 459, 461, 462 that were still ongoing at the time of writing.
Events OGLE-2003-BUL-137 and
earlier were anounced by the EWS in a single email at the beginning of
the 2003 Bulge season and thus not alerted in real time.
OGLE-2003-BUL-238 (A. Gould 2004, private communication)
and 262 \citep{Yoo03} are events in which the lens is known to
have transited the
source and OGLE-2003-BUL-208 and 222 may also involve
finite source effects. These events have not been excluded.
For the remaining
sample of 267 events, I have used only the OGLE III data taken 
before the EWS alert time, defined as the reception of the
alert email by the author.
For the zero point of $\Delta t_{\rm 0}$ for each event, I have used
the definition in \S~2 except for cases in which this has not 
occured before
the EWS alert time in which case the latter has been used 
as
the zero point.

As mentioned in \S~1, different fitting codes can produce
different estimates of maximum magnifications, particularly for
high-magnification events for which blending may be involved. In
particular, there is a concern that a direct comparison of predictions
with the EWS alert predictions may suffer from such differences.  In
order to compare the maximum magnifications predicted by the Bayesian
method with those predicted using $\chi^2$ fitting, I have thus used
very similar computer codes to make $p({\underline \theta}|D,O)$ and
$\chi^2$ optimisations, electing not to use the EWS-fitted parameters.
To avoid the problem of slightly different blending parameters
resulting in large differences in derived magnifications, for each event I
compute the brightness increase, $\Delta m = m_{\rm base} - m(t_{\rm 0})$, 
where
$m(t_{\rm 0})$ is the magnitude at $t_{\rm 0}$. The predicted values of
$\Delta m$ using only the pre-alert data for an event are compared
with the values determined using all the data. When all the data are
available, the parameters derived using $\chi^2$ and
$p({\underline\theta}|D,O)$ are almost always identical.
Exceptions
to this are in a few cases for which there are no data over the peak to 
constrain
the fits.

\subsection{Comparison of Bayesian and $\chi^2$ predictions}

\begin{figure}
\plotone{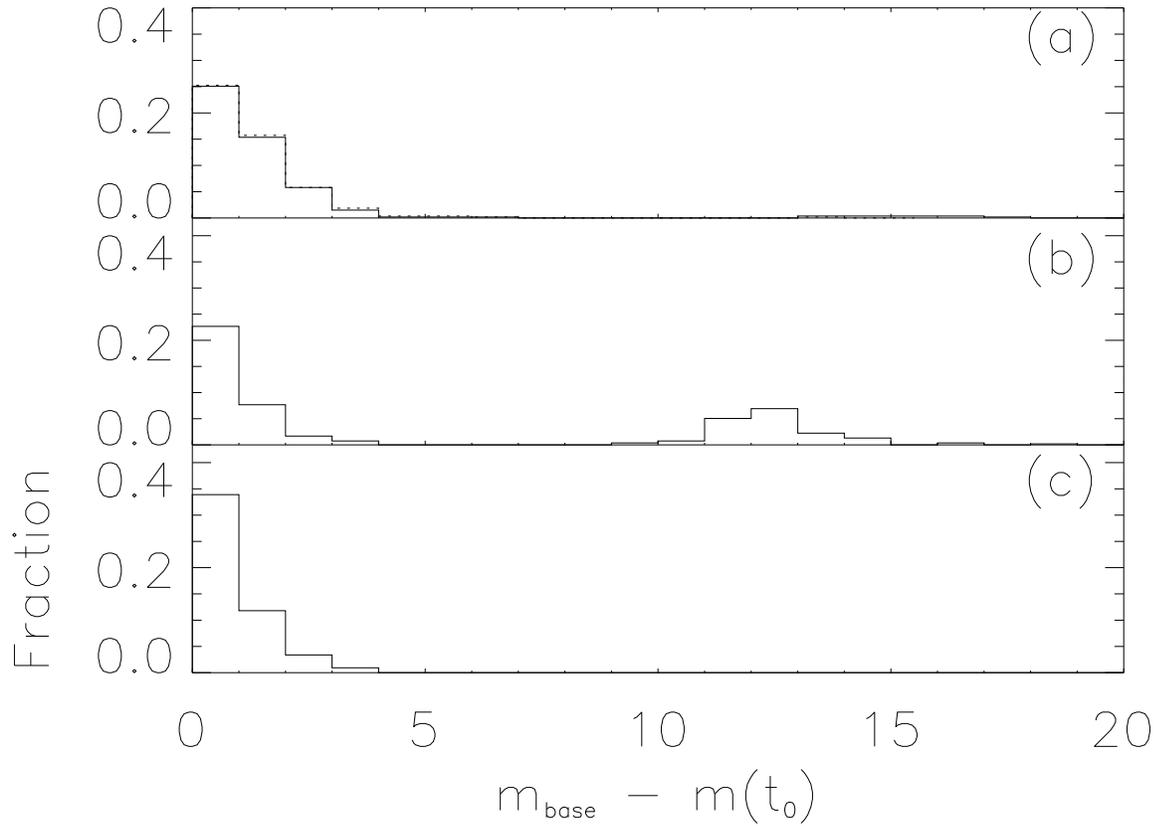}
\caption{Fractional distributions of (a) OGLE peak magnifications with 
         (b) $\chi^{2}$ predictions and (c) Bayesian predictions at
         the time of alert. In panel (a) the solid line represents the
         $\chi^{2}$ fits and the (mostly overplotted) dotted line
         represents the Bayesian fits. \label{fig3}}
\end{figure}

\begin{figure}
\plotone{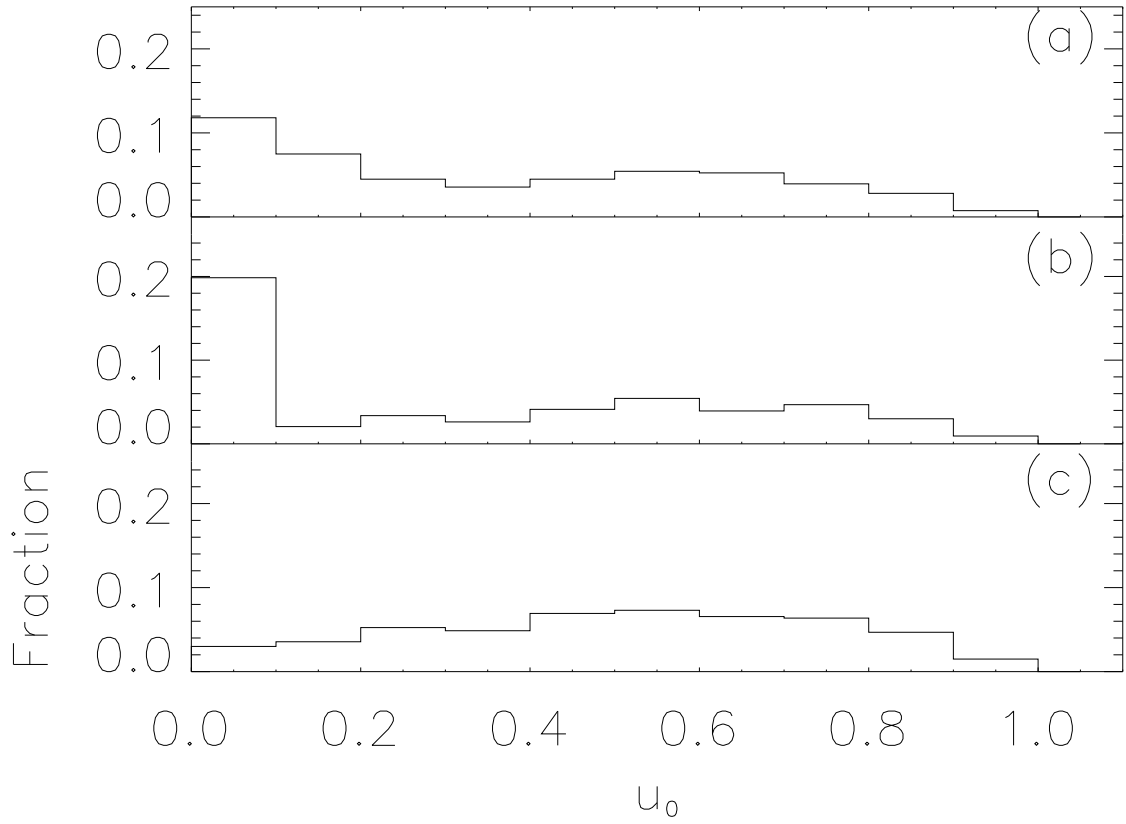}
\caption{Fractional distributions of $u_{\rm 0}$ for the 2003 OGLE 
         PSPL events. Panel (a) shows the values derived using all
         the data while panels (b) and (c) are respectively the 
         $\chi^{2}$ and Bayesian predictions at the time of alert.
         \label{fig4}}
\end{figure}

\begin{figure}
\plotone{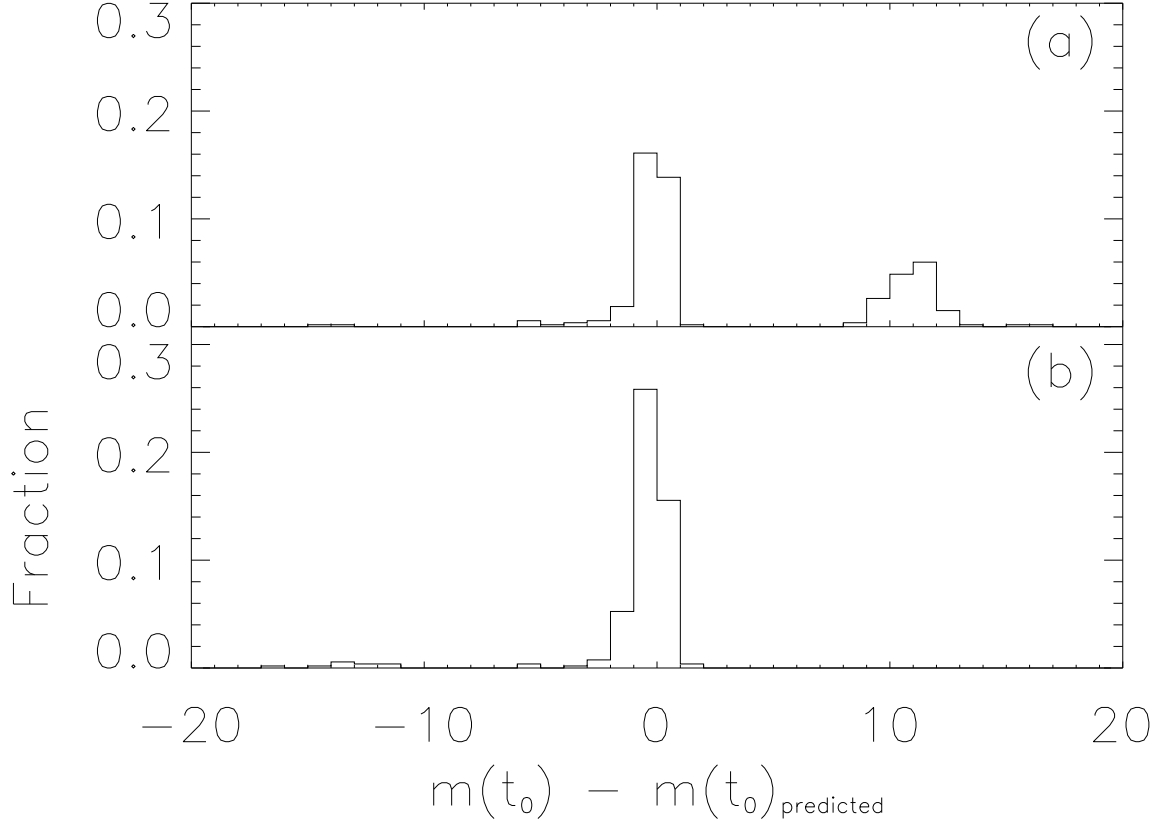}
\caption{Fractional distributions of predicted peak magnitudes minus 
         eventual peak 
         magnitudes for (a) $\chi^{2}$ and (b) Bayesian
         models. \label{fig5}}
\end{figure}

The predictive performances of the Bayesian and $\chi^2$ models at the
time of EWS alert are illustrated in Figures~\ref{fig3} -- \ref{fig5}.
Figure~\ref{fig3} shows the distributions of predicted peak
magnifications for both models and compares these with the eventual
values. Figure~\ref{fig4} shows the same data as a function of $u_{\rm 0}$.
For the $\chi^2$ models, there is clearly a population of low $u_{\rm 0}$
events with predicted brightenings of more than 10 magnitudes that do
not eventuate. Such overpredictions are not present in the Bayesian
fitted models. On the other hand, there is a tendency for the
Bayesian models to underpredict the peak, and at alert time to
fail to predict the small population of high magnification events
in Figure~\ref{fig3}(a).
In Figure~\ref{fig5} I compare the distribution of the differences in 
predicted vs actual
brightenings for both models. Again, the tendency for the $\chi^2$
fits to overestimate the peak is obvious.

\section{Case studies}

As pointed out in \S~4.1, Baysesian
solutions to early lightcurve data often fail to indicate
the nature of high magnification events. It would be of
concern if high magnification events were not observed due
to this tendency. To illustrate in more detail the behavior
of Bayesian vs $\chi^2$ models, I consider here examples
of low and high magnification events. 
These examples 
show several generic aspects of how  Bayesian vs $\chi^2$
solutions evolve as data accumulate.

\begin{figure}
\plotone{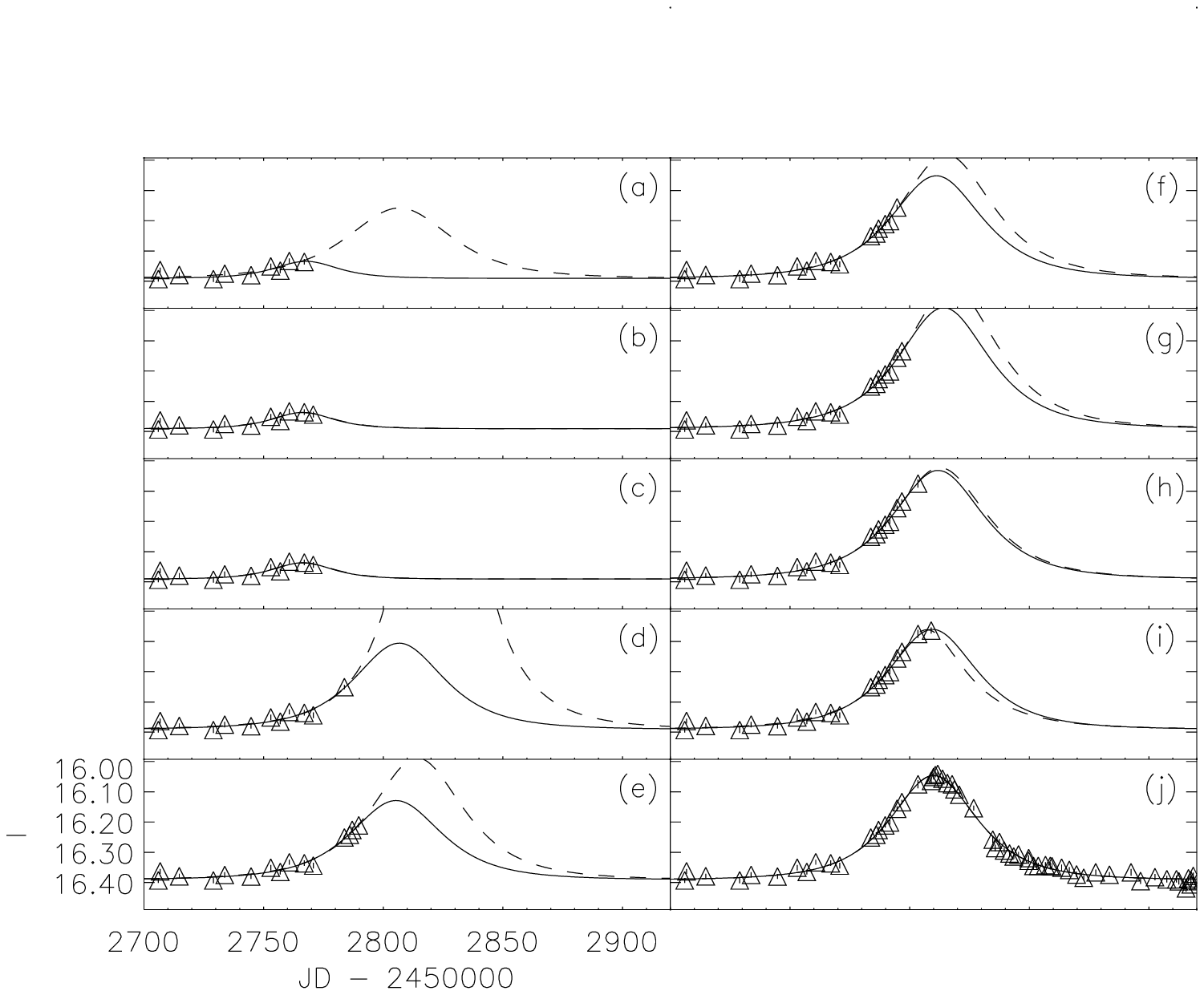}
\caption{Evolution of Bayesian (solid line) and $\chi^{2}$ (dashed line) 
         fits to OGLE-2003-BLG-171. In panels (b), (c), (i) and (j) the 
         dashed line
         is overprinted by the solid line. Axis ranges are the same for all 
         panels.
         \label{fig6}}
\end{figure}

OGLE-2003-BLG-171 was a low magnification event ($A_{\rm 0} =
1.37$). At this magnification, the source star barely passes within
one Einstein radius of the lens and the event is unsuitable for detecting 
a planetary
anomaly. 
This is typical of the type of event that a follow-up program
should avoid observing.
Figure~\ref{fig6} shows $\chi^2$ and Bayesian fitted
lightcurves at 5 day intervals as the event evolves from its alert
date. The predicted maximum magnifications corresponding to
each panel are listed in Table~\ref{tbl1}. 
At alert, the event is predicted to be of low magnification but
by JD 2452785 (panel d in Fig.~\ref{fig6}), the $\chi^2$ solution
suggests a high magnification, albeit with a large uncertainty. 
Since the lightcurve appears to be rising rapidly, follow-up
programs may well begin observing the event in order to improve
on the high uncertainty in the predicted peak magnification. 
As
more data accumulate, the low-magnification nature of the event
becomes apparent. Although the true nature of the event
would be identified relatively quickly by a follow-up observing
program, there is a not-insignificant overhead associated
with adding the event to the program. 
In contrast to the $\chi^2$ fit, the predicted peak
magnification for the Bayesian solution changes steadily with time. 
At no time is a high-magnification event suggested and a follow-up
strategy based on this method would ignore the event.

\begin{figure}
\plotone{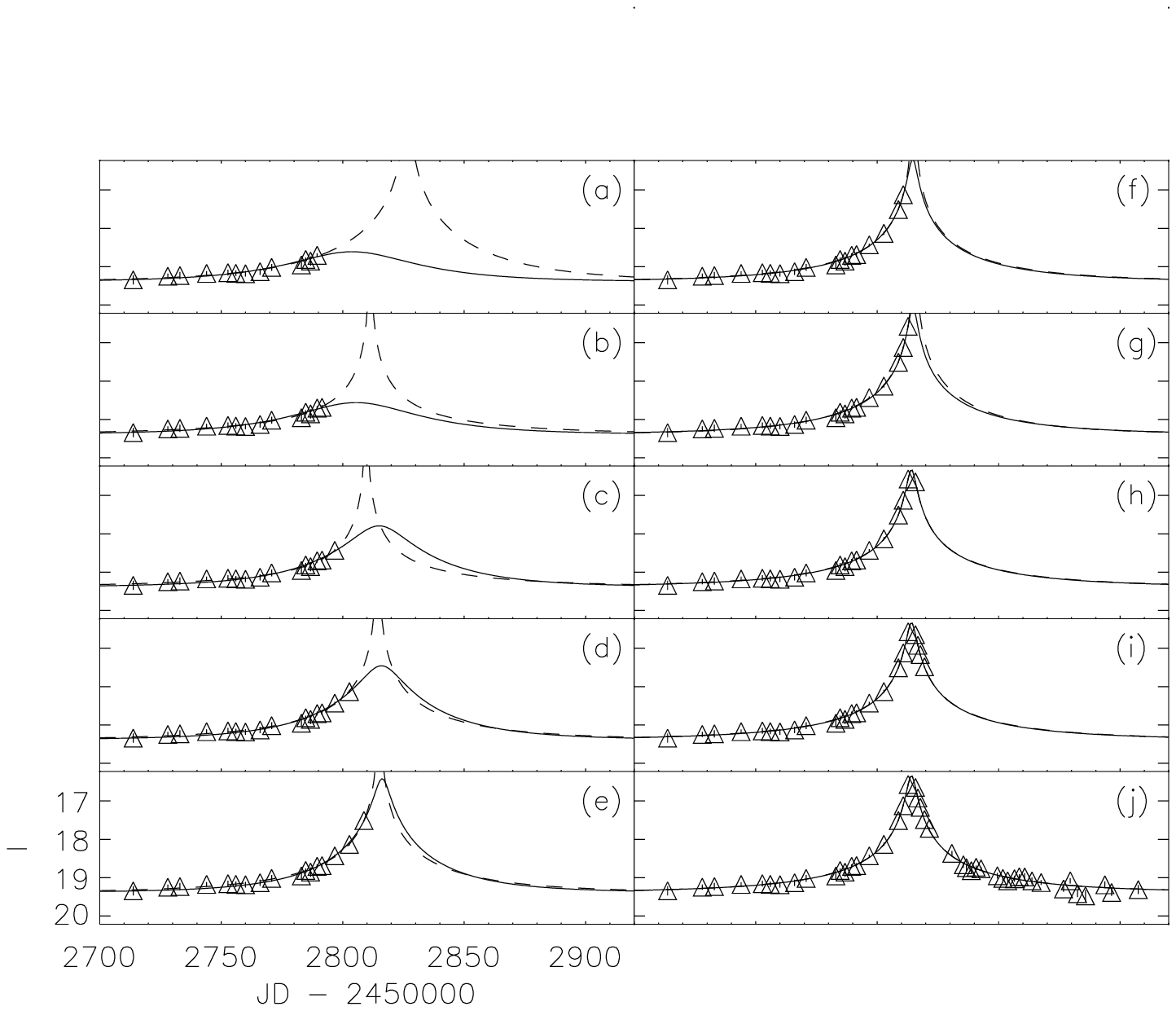}
\caption{Evolution of Bayesian (solid line) and $\chi^{2}$ (dashed line)
         fits to OGLE-2003-BLG-208. In panels (h) -- (j) the dashed line
         is overprinted by the solid line. Axis ranges are the same for all 
         panels.
         \label{fig7}}
\end{figure}

OGLE-2003-BLG-208 (Fig.~\ref{fig7}, Table~\ref{tbl2}) reached a moderately high
magnification ($A_{\rm 0} \approx 45$, $A_{\rm 0,unblended} \approx
17$). The projected source trajectory passed as close as 0.02 $\theta_{\rm
E}$ to the lens and thus had a high probability of intersecting a
central caustic if it were present. The alert date for this event
corresponds to panel (c) in Figure \ref{fig7} at which time
the predictions of peak
magnification are $2 \times 10^5$ and 4.4 for the $\chi^2$ and
Bayesian solutions respectively. 
The Bayesian prediction of $A_{\rm 0} = 4.4$ is sufficiently
high to warrant the
attention of a follow-up observing program such as PLANET.
As data acumulate, the $\chi^2$
predicted peak
magnification rises until reaching $5 \times 10^6$ in panel (g).
The true peak magnification ($A_{\rm 0} \simeq 48$ starts to become 
apparent from  panel (h) as
the event peaks. In contrast, the Bayesian predicted peak magnification
rises steadily until the true peak magnification is identified
from around the time of panels (f) -- (g).

The behavior illustrated by these two examples is typical.  For low
magnification events, the Bayesian model never indicates them as being
worthy of observational follow-up. For high magnification events that
should be observed, the peak magnification is initially underestimated
but adjusts to an appropriate prediction as soon as the data
indicate. In all cases examined, this occurs relatively early in the
event when the magnification, $A \lesssim 3$. For both high and low
magnification events, the Bayesian predicted peak magnification
changes smoothly while the $\chi^2$ prediction is prone to 
large changes as new data points are included. The Bayesian
solutions usually converge to the correct amplification earlier 
than the $\chi^2$ solutions.

\section{Summary}

High magnification events provide the best
opportunity for detecting signals of planets around lens stars
and for obtaining upper limits on their abundances.
Intensive photometric monitoring programs are
hampered currently by difficulties in identifying high magnification
events well before peak. Systems that use $\chi^2$ minimization
to fit PSPL models to early data are prone to exagerated 
predictions of peak magnification. Such prections
induce observers to spend their time monitoring events that
ultimately have little statistical power.

I have shown here that a predictive system based on a Bayesian
formalism that takes account of the characteristics of a
detection system is immune to such behavior. Although such
a Bayesian system tends to initially underpredict the peak 
for high magnification events, accurate prediction occurs as 
soon as sufficient data accumulate to justify the assertion.
In all cases examined, this occurs well ahead of peak in their
associated lightcurves and early enough for the events
to be targeted for observation. Implementation of such a
system based on the OGLE Early Warning System should result
in much improved observing productivity for the 2004 season.

\acknowledgments
I am grateful to Martin Dominik for his comments on an earlier
version of this paper. I think the referee, Andy Gould, for 
his suggested improvements to the manuscript. This work was supported by 
the Marsden Fund under contract UOC302.

\begin{deluxetable}{rrrrrrrrrrrr}
\tabletypesize{\scriptsize}
\tablecolumns{12}
\tablecaption{Peak magnifications for the $\chi^2$ and Bayesian
fits to OGLE-2003-BLG-171 corresponding to the panels in Figure~\ref{fig4}. 
\label{tbl1}}
\tablehead{
\colhead{} & \multicolumn{5}{c}{$\chi^2$} & \colhead{} &
\multicolumn{5}{c}{Bayesian} \\
\cline{2-6} \cline{8-12} \\
\colhead{Panel} & \colhead{$A$} & \colhead{$\sigma_{A}$} & \colhead{$f_{\rm bl}$} & \colhead{$\sigma_{f_{\rm bl}}$} & \colhead{$A_{unblended}$} & \colhead{} &
\colhead{$A$} & \colhead{$\sigma_{A}$} &  \colhead{$f_{\rm bl}$} & \colhead{$\sigma_{f_{\rm bl}}$} & \colhead{$A_{unblended}$} 
}
\startdata
a &     1.240 & 0.670 & 1.000 & --    &     1.240 & &    1.053 & 0.014 & 1.000 & --   & 1.053 \\
b &     1.050 & 0.009 & 1.000 & --    &     1.050 & &    1.051 & 0.009 & 1.000 & --   & 1.051 \\
c &     1.050 & 0.009 & 1.000 & --    &     1.050 & &    1.051 & 0.009 & 1.000 & --   & 1.051 \\
d & 34934.748 & --    & 1.000 & --    & 34934.748 & &    1.300 & 0.314 & 1.000 & --   & 1.300 \\
e &     1.449 & 0.557 & 1.000 & --    &     1.449 & &    1.273 & 0.150 & 1.000 & --   & 1.273 \\
f &     1.459 & 0.330 & 1.000 & --    &     1.459 & &    1.367 & 0.169 & 1.000 & --   & 1.367 \\
g &     1.539 & 0.353 & 1.000 & --    &     1.539 & &    1.446 & 0.199 & 1.000 & --   & 1.446 \\
h &     1.404 & 0.082 & 1.000 & --    &     1.404 & &    1.392 & 0.072 & 1.000 & --   & 1.392 \\
i &     2.524 & 1.366 & 0.233 & 0.206 &     1.355 & &    1.356 & 0.019 & 1.000 & --   & 1.356 \\
j &     1.374 & 0.005 & 1.000 & --    &     1.374 & &    1.374 & 0.005 & 1.000 & --   & 1.374 \\
\enddata
\end{deluxetable}

\begin{deluxetable}{rrrrrrrrrrrr}
\tabletypesize{\scriptsize}
\tablecolumns{12}
\tablecaption{Peak magnifications, blend fractions and deblended peak magnifications for the 
$\chi^2$ and Bayesian
fits to OGLE-2003-BLG-208 corresponding to the panels in Figure~\ref{fig5}. 
\label{tbl2}}
\tablehead{
\colhead{} & \multicolumn{5}{c}{$\chi^2$} & \colhead{} &
\multicolumn{5}{c}{Bayesian} \\
\cline{2-6} \cline{8-12} \\
\colhead{Panel} & \colhead{$A$} & \colhead{$\sigma_{A}$} & \colhead{$f_{\rm bl}$} & \colhead{$\sigma_{f_{\rm bl}}$} & \colhead{$A_{unblended}$} & \colhead{} &
\colhead{$A$} & \colhead{$\sigma_{A}$} &  \colhead{$f_{\rm bl}$} & \colhead{$\sigma_{f_{\rm bl}}$} & \colhead{$A_{unblended}$} 
}
\startdata
a &   63601.415 & --     & 1.000 & --    &   63601.415 & &  2.055 &  0.657 & 1.000 & --    &  2.055 \\
b &  138528.609 & --     & 0.209 & 0.349 &   28987.904 & &  2.150 &  0.700 & 1.000 & --    &  2.150 \\
c &  226215.441 & --     & 0.170 & 0.168 &   38337.850 & &  4.362 &  3.407 & 1.000 & --    &  4.362 \\
d &  224373.536 & --     & 0.382 & 0.324 &   85643.167 & &  5.934 &  4.681 & 1.000 & --    &  5.934 \\
e &  778431.912 & --     & 0.438 & 0.232 &  340873.531 & & 15.425 & 14.683 & 1.000 & --    & 15.425 \\
f & 1669005.679 & --     & 0.412 & 0.159 &  687516.769 & & 41.122 & 31.856 & 0.452 & 0.007 & 19.150 \\
g & 4966165.089 & --     & 0.586 & 0.138 & 2912064.088 & & 46.816 & 12.146 & 0.456 & 0.004 & 21.888 \\
h &      55.641 & 15.221 & 0.293 & 0.076 &      17.022 & & 49.816 &  1.378 & 0.326 & 0.002 & 16.893 \\
i &      53.452 & 13.513 & 0.303 & 0.073 &      16.916 & & 48.446 &  1.356 & 0.333 & 0.002 & 16.813 \\
j &      48.204 &  8.750 & 0.334 & 0.056 &      16.740 & & 45.814 &  1.096 & 0.350 & 0.002 & 16.661 \\
\enddata
\end{deluxetable}


\begin{thebibliography}{}
\bibitem[Albrow et al.(1998)]{Albrow98}
   Albrow, M.D., et al., 1998, \apj, 509, 687
\bibitem[Albrow et al.(2001)]{Albrow01}
   Albrow, M.D., et al., 2001, \apj, 556, 113
\bibitem[Alcock et al.(1993)]{Alcock93}
   Alcock, C., et al., 1993, \nat, 365, 621
\bibitem[Alcock et al.(1995)]{Alcock95}
   Alcock, C., et al., 1995, \apj, 445, 133
\bibitem[Alcock et al.(1997a)]{Alcock97a}
   Alcock, C., et al., 1997a, \apj, 479, 119
\bibitem[Alcock et al.(1997b)]{Alcock97b}
   Alcock, C., et al., 1997b, \apj, 491, 436
\bibitem[Alcock et al.(2000)]{Alcock00}
   Alcock, C., et al., 2000, \apj, 541, 734
\bibitem[Aubourg et al.(1993)]{Aubourg93}
   Aubourg, E., et al., 1993, \nat, 365, 623
\bibitem[Bond et al.(2002)]{Bond02}
   Bond, I.A., et al, 2002, \mnras, 331, 19
\bibitem[Dominik et al.(2002)]{Dominik02}
   Dominik, M., et al, 2002, Planetary and Space Science, 50, 299
\bibitem[Gaudi, Naber \& Sackett(1998)]{Gaudi98}
   Gaudi, B.S., Naber, R.M., Sackett, P.D., 1998, \apj, 500, 33
\bibitem[Gaudi et al.(2002)]{Gaudi02}
   Gaudi, B.S., et al., 2002, \apj, 566, 463
\bibitem[Gould(1994)]{Gould94}
   Gould, A., 1994, \apj, 421, L71
\bibitem[Griest \& Safizadeh(1998)]{Griest98}
   Griest, K., Safizadeh, N., 1998, \apj, 500, 37
\bibitem[Heyrovsk\'y(2003)]{Heyrovsky03}
   Heyrovsk\'y, D., 2003, \apj, 594, 464
\bibitem[Loredo(1990)]{Loredo90}
   Loredo, T.J., 1990, in P.F. Fougere, ed, Maximum Entropy and Bayesian
   Methods, Kluwer, Dordrecht, pp81-142
\bibitem[Paczy{\'n}ski(1986)]{Paczynski86}
   Paczynski, B., 1986, \apj, 304, 1
\bibitem[Rhie et al.(1999)]{Rhie99}
   Rhie, S.H., 1999, \apj, 522, 1037
\bibitem[Sivia(1996)]{Sivia96}
   Sivia, D.S., 1996, Data Analysis. Oxford University Press, Oxford
\bibitem[Udalski et al.(1992)]{Udalski92}
   Udalski, A.,  et al., 1992, Acta Astron., 42, 253
\bibitem[Udalski et al.(1993)]{Udalski93}
   Udalski, A.,  et al., 1993, Acta Astron., 43, 289
\bibitem[Udalski et al.(1994a)]{Udalski94a}
   Udalski, A.,  et al., 1994a, Acta Astron., 44, 165
\bibitem[Udalski et al.(1994b)]{Udalski94b}
   Udalski, A.,  et al., 1994b, Acta Astron., 44, 227
\bibitem[Udalski et al.(2000)]{Udalski00}
   Udalski, A.,  et al., 2000, Acta Astron., 50, 1
\bibitem[Udalski(2003)]{Udalski03}
   Udalski, A., 2003, Acta Astron., 53, 291
\bibitem[Yoo et al.(2004)]{Yoo03}
   Yoo, J.,  et al., 2004, \apj, in press, astro-ph/0309302
\end{thebibliography}
\end{document}